\documentclass[sigplan,screen,authorversion=true]{acmart}
\usepackage[utf8]{inputenc}
\usepackage[T1]{fontenc}

\usepackage{booktabs}
\usepackage{courier}
\usepackage{enumitem}
\usepackage{hyperref}
\usepackage{listings}
\usepackage{lstlinebgrd}
\usepackage{microtype}
\usepackage{numprint}
\usepackage{softdev}
\usepackage{xspace}

\newcommand{\fnc}[1]{\texttt{#1}}

\makeatletter
\newcommand*\Suppressnumber{%
  \lst@AddToHook{OnNewLine}{%
    \let\thelstnumber\relax%
     \advance\c@lstnumber-\@ne\relax%
    }%
}

\newcommand{\memalloc}[1]{\emph{#1}}

\newcommand\narrowwiden{\textsc{NarrowWiden}\xspace}
\newcommand\escperms{\textsc{EscPerms}\xspace}
\newcommand\escinauthentic{\textsc{EscInauthentic}\xspace}
\newcommand\myundef{\textsc{Undef}\xspace}
\newcommand\overlap{\textsc{Overlap}\xspace}

\newcommand\footnoteurl[2]{\footnote{\url{#1}, accessed {#2}}}

\newcommand\tblnarrowwidencapleak{\tiny{NrwWide}}
\newcommand\tblescperms{\tiny{EscPrms}}
\newcommand\tblescunauthentic{\tiny{EscInauth}}
\newcommand\tblundef{\tiny{Undef}}
\newcommand\tbloverlap{\tiny{Overlap}}

\setlist{topsep=0pt, leftmargin=*}

\title{Picking a CHERI Allocator: Security and Performance Considerations}

\author{Jacob Bramley}
\affiliation{%
    \institution{Arm Limited}
    \city{Cambridge}
    \country{United Kingdom}
}
\author{Dejice Jacob}
\affiliation{%
    \institution{University of Glasgow}
    \city{Glasgow}
    \country{United Kingdom}
}
\orcid{0000-0002-4137-0353}
\author{Andrei Lascu}
\affiliation{%
    \institution{King's College London}
    \city{London}
    \country{United Kingdom}
}
\orcid{0000-0002-6623-2053}
\author{Jeremy Singer}
\affiliation{%
    \institution{University of Glasgow}
    \city{Glasgow}
    \country{United Kingdom}
}
\orcid{0000-0001-9462-6802}
\author{Laurence Tratt}
\affiliation{%
    \institution{King's College London}
    \city{London}
    \country{United Kingdom}
}
\orcid{0000-0002-5258-3805}

\thanks{Authors' URLs: %
    D.~Jacob~[\url{https://www.dcs.gla.ac.uk/~jacobd/}],
    A.~Lascu~[\url{https://andreil.eu/}],
    J.~Singer~[\url{https://www.dcs.gla.ac.uk/~jsinger/}],
    L.~Tratt~[\url{https://tratt.net/laurie/}].
}

\setcopyright{rightsretained}
\copyrightyear{2023}
\acmConference[ISMM '23]{Proceedings of the 2023 ACM SIGPLAN International Symposium on Memory Management}{June 18, 2023}{Orlando, FL, USA}
\acmBooktitle{Proceedings of the 2023 ACM SIGPLAN International Symposium on Memory Management (ISMM '23), June 18, 2023, Orlando, FL, USA}
\acmDOI{10.1145/3591195.3595278}

\begin{document}

\begin{abstract}
\noindent
Several open-source memory allocators have been ported to CHERI, a hardware capability
platform. In this paper we examine the security and performance of these
allocators when run under CheriBSD on Arm's prototype Morello platform.
We introduce a number of security attacks and show that all but one allocator
are vulnerable to some of the attacks --- including the default CheriBSD allocator.
We then show that while some forms of allocator performance are meaningful,
comparing the performance of hybrid and pure capability (i.e.~`running in
non-CHERI vs.~running in CHERI modes') allocators does not currently appear to be meaningful.
Although we do not fully understand the
reasons for this, it seems to be at least as much due to factors such as
immature compiler toolchains and prototype hardware as it is due to the effects of
capabilities on performance.
\end{abstract}

\maketitle

\section{Introduction}

CHERI (Capability Hardware Enhanced RISC Instructions) provides and enforces hardware
capabilities that allow programmers to make strong security guarantees
about the memory safety properties of their programs~\cite{watson20cheriintroduction}. However, capabilities are
not magic --- programmers must first decide which memory safety properties they
wish to enforce and then write their software in such a way to enforce those properties.
Mistakes or oversights undermine
the security guarantees that programmers believe their code possesses.

In this paper we study memory allocators (henceforth just ``allocators'')
in the context of CHERI. Apart from some embedded
systems (which preallocate fixed quantities of memory), allocators are
ubiquitous, because they allow us to write programs that are generic over a
variety of memory usage patterns. Allocators' security properties and performance are a fundamental pillar supporting
the security properties and performance of software in general --- any
security flaws and/or performance problems in allocators thus have
significant, widespread, consequences.

In this paper we show that most CHERI allocators are subject to surprisingly simple attacks.
We then show that while some aspects of allocator performance can be meaningfully
compared, it does not currently appear to be meaningful to compare the performance of hybrid
and pure capability (roughly speaking: ``running in
non-CHERI vs.~running in CHERI modes'') allocators.
We analyse some of the likely factors for this latter case,
which suggest that this may be at least as much due to factors such as immature compiler
toolchains and prototype hardware as to the effects of capabilities on performance.
We do not claim that our
work is definitive, though it does suggest two things: that some allocators
undermine the security properties one might reasonably expect from software running on
pure capability CHERI; and that it is currently difficult to reason about the
performance impact of CHERI on software. Our experiments
are repeatable and can be downloaded from
\url{https://archive.org/details/cheri_allocator_ismm}.

This paper is structured as follows. First, we introduce the necessary
background: a brief overview of capabilities and CHERI (\autoref{sec:cheri});
and our running example, a trivial bump allocator (\autoref{sec:bumppointerallocator}).
We then introduce our study: the allocators
under consideration (\autoref{sec:cheriallocators}); our attacks (\autoref{sec:atks});
a partial performance evaluation (\autoref{sec:performance}) and an analysis
of some of the performance discrepancies we uncovered (\autoref{sec:dissection}).

\section{CHERI Overview}
\label{sec:cheri}

In this section, we provide a simple overview of CHERI, and its major concepts.
Since CHERI has developed over a number of years, and is explained across
a range of documentation and papers, some concepts have acquired more than one name,
or names that subtly conflict with mainstream
software development definitions. We use a single name for each concept, sometimes
introducing new names where we believe that makes things clearer.

A \emph{capability} is a token that gives those who bear it \emph{abilities} to
perform certain actions. By restricting who has access to a given capability,
one can enforce security properties (e.g.~``this part of the software can
only read from memory address range X to Y''). Capabilities have a long  history: \cite{levy84capability}
provides a narrative overview of capability architectures, and
may usefully be augmented by more recent work such as~\cite{miller06robust}.
A good first intuition is that CHERI is a modern version of this longstanding
idea, with finer-grained permissions, and adapted to work on recent
processor instruction sets.

We use the term \emph{CHERI} to refer to the `abstract capability machine'
that software can observe: that is, the combination of a capability hardware
instruction set, an ABI (roughly speaking, the interface between userland and kernel
e.g.~\cite{brooks19cheriabi}), a user-facing library that exposes
capability-related functions, and a CHERI-aware language. (e.g.~CHERI C~\cite{watson20chericprogramming},
an adaption, in the sense of both extending and occasionally altering, of C).
Except where we use the name of a different hardware implementation (e.g.~CHERI
RISC-V), we assume the use of Arm's `Morello' hardware, which is a
prototype ARMv8 chip extended with CHERI instructions.

\label{abilities}
Conceptually, a CHERI system starts with a `root' capability that has
the maximum set of abilities. Each
new \emph{child} capability must be derived from one or more \emph{parent}
capabilities. A child capability must have the same, or fewer, abilities than its
parent: put another way, capabilities' abilities monotonically decrease.
An \emph{authentic}\footnote{CHERI calls these
`tagged' or `valid' (and their inauthentic counterparts `untagged' or `invalid').}
capability is one that has been derived from authentic parents according to
CHERI's rules. Attempts to create capabilities that violate CHERI's rules
cause the hardware to produce an \emph{inauthentic} result, guaranteeing
that capabilities cannot be forged.
On Morello and CHERI RISC-V, capabilities behave as if they are 128 bits in
size, but also carry an additional (129th) bit that records the authenticity of each
capability. Software can read, and unset, the authenticity bit, but cannot set it:
only a child capability
derived, correctly, from authentic parent capabilities can itself be authentic.

A capability consists of a memory
\emph{address}\footnote{This portion of a capability does not \emph{have} to store an
address, though typically it does so, and the CHERI API calls it \texttt{address}.
In the context of this paper, since it always stores an address, we stick with this name.},
and its abilities: a set of \emph{permissions} (only a subset of
which we consider in this paper);
and \emph{bounds}, the memory range on which
the capability can operate.

Permissions include the ability to read / write from / to memory.
A \emph{permissions check} is said to be successful if the permission required
for a given operation is provided by a given capability.

A capability's bounds
are from a \emph{low} (inclusive) to a \emph{high} (exclusive) address: when
we refer to a capability's bounds being of `$x$' bytes we mean that
$\textit{high}-\textit{low}=x$. An address is
\emph{in-bounds} for a given capability if it is contained within the
capability's bounds, or \emph{out-of-bounds} otherwise; a capability is
in (or out) of bounds if its address is in (or out) of
bounds\footnote{\cite{woodruff19chericoncentrate} shows why authentic
capabilities can have an out-of-bounds address.}.
A \emph{bounds check} is said to be successful if a given capability, address, or
address range, is in-bounds for a given capability.

A processor instruction that operates on a capability requires at least one of: an authenticity
check, a permissions check, or a bounds check. If a capability fails the
relevant checks, then either: the hardware produces a
\lstinline{SIGPROT} exception (similar to \lstinline{SIGSEGV})
that terminates the program; or produces an inauthentic capability (where this is not
in violation of the CHERI rules).

CHERI allows both double-width capabilities and
single-width addresses-as-pointers to exist alongside each other at any time.
Conventionally, a program which uses both traditional addresses and
capabilities is said to be operating in \emph{hybrid} mode while a program
which uses only capabilities is in \emph{pure capability} -- henceforth ``purecap'' -- mode.
Some caution is necessary with these terms, because different parts of a system
may be hybrid or purecap. For example, kernels often use hybrid mode while
some parts of userland may use purecap: at least one of the sides in such
a relationship must translate between the hybrid and purecap worlds as necessary.
In this paper we make two simplifying assumptions: that we only have to
consider userland, and that userland is entirely purecap or entirely hybrid.

CHERI does not presuppose a particular Operating System (OS). While there is a
CHERI Linux port, at the time of writing the most mature OS for CHERI hardware
is CheriBSD, a FreeBSD descendent. In this paper we use CheriBSD.

\section{A Basic Pure Capability Allocator}
\label{sec:bumppointerallocator}

To illustrate how CHERI affects allocators, in this section we adapt a simple
non-CHERI aware allocator to become CHERI aware. For simplicity's sake,
we assume that this means at least running successfully on, and preferably
taking advantage of, purecap CHERI.

\begin{figure}[t]
\begin{lstlisting}[
  language=C,
  xleftmargin=1.3em,
  caption={
    A simple, but complete, non-CHERI aware, bump pointer allocator:
    \fnc{malloc} works as per normal; \fnc{free} is a no-op; and
    \fnc{realloc} always allocates a new chunk, copying
    over the old block.
    \fnc{\_\_builtin\_align\_up(v, a)} is an LLVM / clang primitive which rounds
    \texttt{v} up to the next smallest multiple of \texttt{a};
    \texttt{\_Alignof(max\_align\_t)} returns an alignment sufficiently large
    for any scalar type (i.e.~integers and pointers).
  },
label=lst:bump_alloc1]
#include <string.h>
#include <sys/mman.h>

char *heap = NULL;
char *heap_start = NULL;
size_t HEAP_SIZE = 0x1000000000;

void *malloc_init() {
  heap = heap_start = mmap(NULL, HEAP_SIZE,
    PROT_READ | PROT_WRITE,
    MAP_PRIVATE | MAP_ANON, -1, 0);
  return heap;
}

void *malloc(size_t size) {
  if (heap == NULL && !malloc_init())
    return NULL;
  size = __builtin_align_up(size,
    _Alignof(max_align_t));
  if (heap + size >
      heap_start + HEAP_SIZE)
    return NULL;
  heap += size;
  return heap - size;
}

void free(void *ptr) { }

void *realloc(void *ptr, size_t size) {
  void *new_ptr = malloc(size);
  if (new_ptr == NULL) return NULL;
  if (ptr) memcpy(new_ptr, ptr, size);
  return new_ptr;
}
\end{lstlisting}
\end{figure}

\autoref{lst:bump_alloc1} shows a simple, complete, example of a C bump allocator:
\fnc{malloc} works as per normal; \fnc{free} is
a no-op; and \fnc{realloc} always allocates a new chunk of memory. The
allocator reserves a large chunk of memory using a single \fnc{mmap} call then
doles out chunks on each \fnc{malloc} / \fnc{realloc} calls. The bump pointer
moves through the \fnc{mmap}ed chunk until it reaches the upper limit, at which
point the allocator returns \texttt{NULL}. Though intentionally simplistic,
\fnc{realloc} is correct even when the block is increased in size.

\subsection{Adapting the Allocator to CHERI}
\label{sec:adapting_to_cheri}

Perhaps surprisingly, our simple bump allocator compiles, and \fnc{malloc} runs
correctly, on a purecap CHERI system. As this suggests, CHERI C is largely source
compatible with normal C code, though pointer types are transparently `upgraded' to become
\emph{capability types} (on Morello occupying exactly twice the space of a
non-capability pointer). CHERI also implies changes in libraries: on CheriBSD,
for example, \fnc{mmap} returns a capability whose bounds are at least those of
the size requested: from that capability our bump allocator derives new
capabilities that differ in their address but not their bounds. In other
words, calling \fnc{malloc} just once gives the caller the ability to read and
write from all past and future blocks returned by \fnc{malloc}!

\begin{figure}[t]
\begin{lstlisting}[language=C,
  xleftmargin=1.3em,
  caption={
    Replacing the non-CHERI aware \fnc{malloc} from ~\autoref{lst:bump_alloc1}
    with a CHERI-aware alternative
    using the idioms suggested in~{\cite[p.~30]{watson20chericprogramming}}.
    This \fnc{malloc} returns a capability whose bounds are sufficient
    to cover \texttt{size} bytes starting at the capability's address
    (calculated in lines 5--10),
    such that two callers to \fnc{malloc} cannot read or write from
    another block. We also have to update \fnc{realloc} so that it never tries to copy
    more data from the old block than the \texttt{ptr} capability gives
    it access to.
  },
    label=lst:bump_alloc2]
void *malloc(size_t size) {
  if (heap == NULL && !malloc_init())
    return NULL;

  char *new_ptr = __builtin_align_up(
    heap,
    -cheri_representable_alignment_mask(
      size));
  size_t bounds =
    cheri_representable_length(size);
  size_t size_on_heap =
    __builtin_align_up(
      size, _Alignof(max_align_t));

  if (new_ptr + size_on_heap >
      heap_start + HEAP_SIZE)
    return NULL;
  heap = new_ptr + size_on_heap;
  return cheri_bounds_set_exact(
    new_ptr, bounds);
}

void *realloc(void *ptr, size_t size) {
  void *new_ptr = malloc(size);
  if (new_ptr == NULL) return NULL;
  if (ptr)
    memcpy(new_ptr, ptr,
      cheri_length_get(ptr) < size
      ? cheri_length_get(ptr) : size);
  return new_ptr;
}
\end{lstlisting}
\end{figure}

As this suggests, using CHERI without careful thought may give
no additional security benefits. This then raises the question:
how should a secure `CHERI aware' allocator behave? There can be no single
answer to this question, but we believe that most programmers would at least expect
\texttt{malloc} to return a capability whose bounds are restricted to the block of memory
allocated. \autoref{lst:bump_alloc2}
shows how to adapt \texttt{malloc} to do this.

The code to create the capability (using the idioms suggested
in~\cite[p.~30]{watson20chericprogramming}) is more involved than one might
first expect. The underlying cause is that there aren't, and cannot reasonably
be, enough bits in CHERI's bounds to precisely represent every possible address
and size. Modern CHERI therefore uses an encoding for bounds that allows small
bounds to be precisely represented, at the expense of larger bounds becoming
progressively less precise~\cite{woodruff19chericoncentrate}. On Morello,
the smallest bound that cannot be precisely represented is 16,385 bytes,
which is rounded up to 16,392 bytes\footnote{For CHERI RISC-V the first unrepresentable length is
4,097 bytes, which is rounded up to 4,104.}. Our capability aware
\texttt{malloc} thus has to ensure that both the capability's low and high
bound addresses are rounded down and up (respectively) in a way that ensures
that the address and size can be fully covered.

The two versions of our allocator have meaningfully different security properties,
even when we run both on a purecap system. For example, consider this simple
C snippet which models a buffer overrun:

\begin{lstlisting}[language=C]
char *b = malloc(1);
b[0] = 'a';
b[1] = 'b';
\end{lstlisting}

\noindent
On a non-CHERI system, or a CHERI system with \autoref{lst:bump_alloc1}
as an allocator, this snippet compiles and runs without error. However, using
the allocator from \autoref{lst:bump_alloc2} on a purecap CHERI system,
while the snippet compiles, the tighter
capability bounds returned by \texttt{malloc} cause a \lstinline{SIGPROT}
when executing line 3. This demonstrates how CHERI can prevent
programmer errors becoming security violations.

However, just because a program compiles with CHERI C does not guarantee that
it will run without issue: when run on CHERI, the \fnc{realloc} in~\autoref{lst:bump_alloc1}
causes a \lstinline{SIGPROT} when asked to
increase the size of a block. This occurs because \fnc{memcpy} tries to copy
beyond the bounds of the input capability (e.g.~if the existing block is 8
bytes and we ask to resize it to 16 bytes, \fnc{memcpy} tries to read
16 bytes from a capability whose bounds are 8 bytes). On a non-CHERI system,
this is not a security violation, but it is treated as one on CHERI. In
\autoref{lst:bump_alloc2} we thus provide an updated \fnc{realloc} which
uses \texttt{cheri\_length\_get} (which returns a capability's bounds
in bytes) to ensure that it never copies more data than the input capability's
bounds allow.

\section{CHERI Allocators}
\label{sec:cheriallocators}

In this paper we consider a number of allocators that are available for
CheriBSD. We first explain the set of allocators we use, before
exploring in more detail how the allocators have been adapted (if at all) for
CHERI.

\subsection{The Allocators Under Consideration}

\begin{table}[tb]
\begin{center}
\begin{tabular}{llrrr}
\toprule
Allocator & Version & SLoC & \multicolumn{2}{c}{Changed}\\
\cmidrule(lr){4-5}
  &   &   & LoC & \multicolumn{1}{c}{\%}\\
\midrule
bump-alloc & 21cb5f38 & \numprint{61} & \numprint{31} & 50.81\\
dlmalloc-cheribuild & 9cfbb169 & \numprint{3475} & \numprint{231} & 6.65\\
jemalloc & cc4e4c05 & \numprint{28755} & \numprint{116} & 0.40\\
libmalloc-simple & 62175107 & \numprint{408} & \numprint{43} & 10.54\\
snmalloc-cheribuild & 888d182b & \numprint{14669} & \numprint{180} & 1.23\\
snmalloc-repo & 0a5eb403 & \numprint{21342} & \numprint{212} & 0.99 \\
\midrule
dlmalloc-pkg64c & 2.8.6 & - & - & -\\
ptmalloc & 3.0\_2 & - & - & -

\\ \bottomrule
\end{tabular}
\end{center}
  \caption{The allocators we examined, their size in Source Lines of Code
  (SLoC), and the number of lines changed (as an absolute value and relative
  percentage) to adapt them for purecap CheriBSD. The top portion
  of the table shows the allocators which
  passed a basic test and are used in our experiments; the bottom portion
  shows the allocators which failed a basic test.}
\label{tab:allocator_summary}
\end{table}

A number of allocators are available for purecap CheriBSD, installable via three
different routes: as part of the base distribution; via CheriBSD
\emph{packages}; or via external sources.  We examined allocators
available from all three sources. We excluded
allocators aimed at debugging (e.g.~\emph{ElectricFence}).
We then ran a simple validation test, \fnc{malloc}ing a block of memory,
copying data into the block, and then \fnc{free}ing the block: we
excluded any allocator which failed this test.

On that basis, the allocators we consider in this paper, and the names we use
for them for the rest of this paper, are as follows:

\setlength{\leftskip}{6pt}

\vspace{6pt}\noindent
\memalloc{jemalloc}, a modified version of the well-known
    allocator~\cite{evans06scalable}: this is the default allocator
    for CheriBSD.

\vspace{6pt}\noindent
\memalloc{libmalloc-simple}\footnoteurl{https://github.com/CTSRD-CHERI/cheribsd/commit/e85ccde6d78d40f130ebf126a001589d75d60473}{23rd
    February 2023}, a port of the allocator in the FreeBSD utility
    \texttt{rtld-elf}\footnoteurl{https://github.com/freebsd/freebsd-src/blob/releng/4.3/libexec/rtld-elf/malloc.c}{
    23rd of February 2023}, based on Kingsley's malloc from 4.2BSD.

\vspace{6pt}\noindent
\memalloc{snmalloc-cheribuild}, a version of
    \emph{snmalloc}~\cite{lietar19snmalloc} that can be installed via
    \texttt{cheribuild}. We found this version to have several problems
    which we rectified by manually building a newer version from \emph{snmalloc}'s
    GitHub repository. We term this more recent version \texttt{snmalloc-repo}.

\vspace{6pt}\noindent
\memalloc{dlmalloc-cheribuild}, a modified version of the well-known
    allocator~\cite{lea96memory},
    installable via \texttt{cheribuild}. \memalloc{dlmalloc-pkg64c} is an
    unmodified version of the allocator, available as a package in CheriBSD. Both
    versions are based on dlmalloc 2.8.6.

\vspace{6pt}\noindent
\memalloc{ptmalloc}~\cite{gloger06ptmalloc} is an extension of
    \memalloc{dlmalloc}, with added support for multiple threads.

\vspace{6pt}\noindent
\memalloc{bump-alloc-nocheri} is the simple, non-CHERI-aware bump
    allocator from~\autoref{lst:bump_alloc1}. Conversely, the CHERI aware version is
    \memalloc{bump-alloc-cheri},  presented
    in~\autoref{lst:bump_alloc2}.

\setlength{\leftskip}{0pt}

\vspace{6pt}\noindent
\autoref{tab:allocator_summary} shows the version of each allocator we used.
We have not included two other
major memory allocators that have only been
partly ported to CHERI: the \emph{Boehm-Demers-Weiser} conservative garbage
collector; and the \emph{WebKit} garbage collector.

\subsection{How Much Have the Allocators Been Adapted for CHERI?}
\label{sec:rqs}

As we saw from \autoref{lst:bump_alloc1}, simple allocators may not need
adapting for CHERI, though they are then likely to derive only minor security gains. In
practice, we expect most allocators to incorporate at least the capability bounds enforcement
of \autoref{lst:bump_alloc2}. Indeed, more sophisticated allocators tend
to crash on CHERI without at least some modifications. For example, most of the allocators available
via CheriBSD's package installer (e.g.~dlmalloc-pkg64c) have had no source-level
changes for CHERI: they compile correctly but crash on even the most trivial examples.

Understanding the details of the CHERI modifications to all of the allocators under
consideration is beyond the scope of this work.
Instead, \autoref{tab:allocator_summary} shows what proportion of an
allocator's LoC are `CHERI specific' by calculating the percentage of lines of code contained between
\texttt{\#ifdef CHERI} blocks and similarly guarded code. This count is an under-approximation, as some
code outside such \texttt{\#ifdef} blocks may also have been adapted, but it
gives a rough idea of the extent of changes.

With the exception of the extremely small
\texttt{bump-alloc} and \texttt{libmalloc-simple},
the pure capability memory manager libraries in Table~\ref{tab:allocator_summary} have a mean
2.31\% of their SLoC changed.
Although this is a relatively small portion,
it is two orders of magnitude larger than the
0.026\% lines that were adapted when porting a desktop environment (including X11 and
KDE)~\cite{watson21assessing}. It is a reasonable assumption that the
lower-level, and more platform dependent, nature of allocators
requires more LoC to be adapted.

\section{The Attacks}
\label{sec:atks}

\begin{table}[t]
\begin{center}
\begin{tabular}{lccccc}
\toprule
Allocator & \tblescunauthentic & \tblescperms & \tblnarrowwidencapleak & \tbloverlap & \tblundef\\
\midrule
bump-alloc-cheri & $\checkmark$ & $\times$ & $\checkmark$ & $\checkmark$ & $\checkmark$\\
bump-alloc-nocheri & $\checkmark$ & $\oslash$ & $\checkmark$ & $\times$ & $\checkmark$\\
dlmalloc-cheribuild & $\checkmark$ & $\times$ & $\times$ & $\checkmark$ & $\checkmark$\\
jemalloc & $\checkmark$ & $\times$ & $\times$ & $\checkmark$ & $\times$\\
libmalloc-simple & $\checkmark$ & $\checkmark$ & $\times$ & $\checkmark$ & $\times$\\
snmalloc-cheribuild & $\checkmark$ & $\checkmark$ & $\checkmark$ & $\oslash$ & $\checkmark$\\
snmalloc-repo & $\checkmark$ & $\checkmark$ & $\checkmark$ & $\checkmark$ & $\checkmark$

\\ \bottomrule
\end{tabular}
\caption{Attacks per allocator: $\times$ indicates that an allocator is vulnerable to an attack;
  $\checkmark$ that the allocator is invulnerable; and $\oslash$ a failure for other
  reasons (e.g.~a segfault).}
\label{tab:atks}
\end{center}
\end{table}

Our definition of CHERI in~\autoref{sec:cheri} might suggest that software
running on CHERI hardware is invulnerable to attack. Rather, CHERI gives us the
tools to make secure software, but it is up to us to use them correctly --- and wisely.
We must decide which attack model is relevant to our use-case, and then write,
or adjust, the software, to withstand such attacks. In our context, allocators are
subject to spatial (e.g.~buffer overrun) or temporal (e.g.~a sequence of
function calls) attacks, and those attacks can either target an allocators'
internals (e.g.~corrupting private data-structures) or its interface
(e.g.~allowing user code to bypass security checks).

In this section we introduce a number of simple `attacks' on CHERI allocators (4
temporal and 1 spatial) and then run those attacks on the allocators from
\autoref{sec:cheriallocators}, with the results shown in \autoref{tab:atks}.
Even the default CheriBSD allocator is vulnerable to some attacks: only
snmalloc is invulnerable.

In the rest of this section we explain each attack, giving C code using the
CHERI API. Our code examples assume that we start with a
`non-attacker' who allocates memory and hands it over to another part of the system which has been
taken over by an `\colorbox{gray!20}{attacker}' (whose code has a light grey background).
For each (allocator, attack) pair, we state whether it is
vulnerable, invulnerable, or whether the attack fails for other reasons.
We model this via a series of \lstinline{assert}s:
if all the \lstinline{assert}s pass, the attack is successful. The code we show
in the paper is elided relative to the version we run, which contains changes
that makes it possible for us to automate the running of the attacks over
multiple allocators. The
full code (available as part of our experiment) must be considered the definitive source of
truth for \autoref{sec:cheriallocators}.

Our descriptions use the following CHERI functions:

\setlength{\leftskip}{6pt}

\vspace{6pt}
\noindent\texttt{void *cheri\_address\_set(void *c, \\ptraddr\_t a)}
\hspace{\parindent}Takes a capability \texttt{c} as input and returns a capability
that is a copy of \texttt{c} except with the address \texttt{a}.
\texttt{ptraddr\_t} is a CHERI C integer type that is guaranteed to be big enough
to represent addresses but, unlike \texttt{intptr\_t}, is not big enough to
represent capabilities.

\vspace{6pt}
\noindent\texttt{ptraddr\_t cheri\_base\_get(void *c)}\\
Returns the address of the lower bound of a capability \texttt{c}.

\vspace{6pt}
\noindent\texttt{void *cheri\_bounds\_set(void *c, \\size\_t s)}
\hspace{\parindent}Takes a capability \texttt{c} as input and returns a new capability
that is a copy of \texttt{c} except with bounds \texttt{s}.

\vspace{6pt}
\noindent\texttt{size\_t cheri\_length\_get(void *c)}\\
Returns the bounds of a capability \texttt{c}.

\vspace{6pt}
\noindent\texttt{\_Bool cheri\_tag\_get(void *c)}\\
Returns true if the capability \texttt{c} is authentic.

\vspace{6pt}
\noindent\texttt{void* cheri\_perms\_and(void *c,\\ size\_t perms)}
\hspace{\parindent}Returns the capability \texttt{c} with its permissions bitwise-ANDed
with \texttt{perms}.

\vspace{6pt}
\noindent\texttt{size\_t cheri\_perms\_get(void *c)}\\
Returns the permissions of capability \texttt{c}.

\setlength{\leftskip}{0pt}

\subsection{\narrowwiden: Narrowing then Widening Can Allow Access to Hidden Data}
\label{narrowwiden}

In the simple bump allocator of \autoref{lst:bump_alloc1}, \texttt{realloc} always
allocates a new block of memory. While this is always correct, it is inefficient,
in part because it requires copying part of the block's existing content.
Most allocators thus try to avoid allocating a new block of memory if: the
new size is the same as, or smaller than, the existing size; or if
the new size would not lead to the block overwriting its nearest neighbour.
The latter optimisation is dangerous for a CHERI allocator.

Consider the case where \texttt{realloc} wants to increase a
block in size, and there is sufficient room to do so without moving the block.
\texttt{realloc} needs to return a capability whose bounds encompass the new
(larger) size. However, such a capability cannot be derived from the input
capability, as doing so would lead to an inauthentic capability, and
we would violate the property that a capability's abilities must monotonically
decrease. Thus, the allocator needs access to a `super'
capability which it can use to derive a capability representing the new bounds.
Let us call the `super' capability \texttt{SC} and introduce a function
\texttt{size\_of\_bucket} which tells us the maximum space available for
the block starting at \texttt{ptr}. Eliding extraneous details (e.g.~about
alignment), \fnc{realloc} will then look as follows:

\begin{lstlisting}[language=C]
void *realloc(void *ptr, size_t size) {
  if (size <= size_of_bucket(ptr)) {
    // No need to reallocate.
    void *new_ptr =
      cheri_address_set(SC, ptr),
    return cheri_bounds_set(
      new_ptr, size);
  } else {
    // Allocate a larger region of memory
    // and copy the old contents.
  }
}
\end{lstlisting}

\noindent Lines 4--7 need to deal with the
case where the block is to be increased in size but will still fit
in its current bucket. We
first use \fnc{cheri\_address\_set} to derive a new capability from
\texttt{SC} whose address is the same as \texttt{ptr} but whose bounds will be
those of \texttt{SC} (lines 5 and 6) before narrowing those bounds to
\texttt{size} (lines 6 and 7).

When implemented in this style, an allocator can be subject to the following
attack:

\begin{lstlisting}[language=C,linebackgroundcolor={\ifnum\value{lstnumber}>4\color{gray!20}\fi}]
uint8_t *arr = malloc(256);
for (uint8_t i = 0; i < 256; i++)
  arr[i] = i;
arr = realloc(arr, 1);
arr = realloc(arr, 256);
for (uint8_t i = 0; i < 256; i++)
  assert(arr[i] == i);
\end{lstlisting}

\noindent
We first allocate a block of memory, receiving a capability with
a bounds of 256 bytes (line 1). We fill the block up with data (lines 2 and 3)
then \fnc{realloc} the block down to a single byte, receiving back a capability
whose bounds are 1 byte\footnote{Some allocators return bounds bigger than 1 byte, though
none we tested returned a bounds of 256 bytes or more.} (line 4).

At this point, we expect to have permanently lost access to the values written to bytes 2-255
in lines 2 and 3 --- if an attacker \fnc{realloc}s the block back to
its original size they should not have access to the values written to bytes
2-255. However, allocators using the optimisation above will often
return a capability that covers the same portion of memory as the original
block, without zeroing it, allowing an attacker to read the original bytes out
unchanged (lines 6 and 7).

It might seem merely undesirable for \fnc{realloc} to allow an attacker access to the
original data, but in a capability system this attack is particularly
egregious if that data contains capabilities, since an attacker can read those
and gain new abilities.

\subsubsection{Mitigations}

Mitigating this attack is relatively simple. When \fnc{realloc} shrinks a
block, any excess storage should be zeroed. Note that we consider this safer
than the seemingly similar alternative of zeroing excess storage when
\fnc{realloc} enlarges a block, because that implies a delay in zeroing that
might give an attacker other unexpected opportunities to read data.

\subsection{\escperms: Escalate Permissions}

When an allocator uses a `super' capability (as seen in
\autoref{narrowwiden}), there may be potential to upgrade a
capability's permissions as shown in the following simple attack:

\begin{lstlisting}[language=C,linebackgroundcolor={\ifnum\value{lstnumber}>6\color{gray!20}\fi}]
uint8_t *arr = malloc(16);
assert(cheri_perms_get(arr)
  & CHERI_PERM_STORE));
arr = cheri_perms_and(arr, 0);
assert((cheri_perms_get(arr)
  & CHERI_PERM_STORE) == 0);
arr = realloc(arr, 16);
assert(cheri_perms_get(arr)
  & CHERI_PERM_STORE);
\end{lstlisting}

\noindent
We first allocate a block and check that the capability returned
is allowed to store data (\texttt{CHERI\_PERM\-\_STORE}) to that block (lines 1--3).
We deliberately remove the store permission (line 4), checking
that this permission really has been removed (lines 5 and 6). We then call
\fnc{realloc} (without changing the block's size) and check whether we have
regained the ability to store data via the capability.

\subsubsection{Mitigations}

There are two ways of mitigating such an attack. The simplest is to AND
the output capability's permissions with the input capability's permissions:
doing so guarantees that the output capability has no more permissions
than the input capability.

However, in some cases, one should consider validating the input capability to
decide whether any action should be possible. For example, if handed a
capability whose address is genuinely an allocated block, but where the
capability has the read and write permissions unset, should \fnc{realloc}
refuse to reallocate the block? Perhaps \fnc{realloc} should check that
the capability handed to it has exactly the same permissions as the
capability handed out by the most recent \fnc{malloc} or \fnc{realloc}?
There are no universal answers, but some cases may be
easier to rule upon than others.

\subsection{\escinauthentic: Escalate Inauthentic Capabilities}
\label{sec:escinauthentic}

An important variant on \escperms is to see whether an allocator will
reallocate a block pointed to by an inauthentic capability and return
an authentic capability:

\begin{lstlisting}[language=C,linebackgroundcolor={\ifnum\value{lstnumber}>4\color{gray!20}\fi}]
uint8_t *arr = malloc(16);
assert(cheri_tag_get(arr));
arr = cheri_tag_clear(arr);
assert(!cheri_tag_get(arr));
arr = realloc(arr, 16);
assert(cheri_tag_get(arr));
\end{lstlisting}

\noindent
Interestingly, none of the allocators we examined was vulnerable to this
attack. However, several fall into something of a
grey zone: despite not explicitly checking the capability's authenticity, the
allocators cause a \lstinline{SIGPROT} when they try to perform an operation on
the inauthentic capability. It is difficult to know whether this was an
expected outcome or not, in the sense that programs often explicitly rely on
null pointer dereferencing causing \texttt{SIGSEGV} to maintain certain
security properties. However, since the outcome is a reasonable one, we have
chosen to give the allocators the benefit of doubt in this case, and have
classified them as
invulnerable to this attack.

\subsection{\myundef: Authentic capabilities from Undefined Behaviour}

It is easy to assume that authentic capabilities can only be derived if one
follows CHERI-C's rules correctly. However, it is possible for an attacker to use undefined
behaviour at the language level to trick an allocator into returning authentic capabilities
that it should not possess as shown in this attack:

\begin{lstlisting}[language=C,linebackgroundcolor={\ifnum\value{lstnumber}>5\color{gray!20}\fi}]
uint8_t *arr = malloc(256);
for (uint8_t i = 0; i < 256; i++)
  arr[i] = i;
arr = realloc(arr, 1);
free(arr);
arr = malloc(256);
for (uint8_t i = 0; i < 256; i++)
  assert(arr[i] == i);
\end{lstlisting}

\noindent
This follows a similar pattern to \narrowwiden. We first allocate a block
and fill it with data (lines 1--3). Although not strictly necessary to demonstrate
the attack, we then reallocate the block down to a single byte, modelling the
case where we pass a capability with few abilities to an attacker (line 4). The
attacker then frees that block (line 5) and immediately allocates a block of the same
size (line 6) hoping that the new block is allocated in the same place as the old
block. If that is the case, they will obtain a capability spanning the same memory as the old block,
which allows them access to secret data (lines 7 and 8).

Interestingly, this attack places both the `non-attack' and `attack' portions
into undefined behaviour. Most obviously, the attack portion of the code reads
data via a capability / pointer that it cannot ensure has been initialised. Less
obviously, both attacker and non-attacker have an equivalent capability (with
the same address and bounds) but, due to capability / pointer provenance rules,
the non-attackers version of the capability is, technically speaking, no
longer valid by those rules. This outcome is unlikely to trouble an attacker.

\subsubsection{Mitigations}

There are no general mitigations for \myundef. For the particular concrete
example, a partial mitigation is for \fnc{free} to scrub memory so that, at
least, whatever was present in the buffer cannot be read by the attacker.
More generally, attackers may sometimes be able to use undefined behaviour to `alias'
a capability. In most such cases, the only solution is to
scan memory looking for all references to a capability whose bounds
encompass an address and render them inauthentic (so-called `revocation'~\cite{xia19cherivoke}),
downgrading a security leak into a denial-of-service.

\subsection{\overlap: Capabilities Whose Memory Bounds Overlap with Another's}
\label{sec:overlap}

A capability's bounds span a portion of memory from a low to a high address. If
an allocator returns two distinct capabilities whose bounds overlap
(e.g.~because of the bounds imprecision we saw in
\autoref{sec:adapting_to_cheri} as a result of
\cite{woodruff19chericoncentrate}), an attacker might be able to read or write
memory they should not have access to. An example attack in this mould is:

\begin{lstlisting}[language=C,linebackgroundcolor={\ifnum\value{lstnumber}>1\color{gray!20}\fi}]
void *b1 = malloc(16);
void *b2 = malloc(16);
assert(
  cheri_base_get(b1) >= cheri_base_get(b2)
  && cheri_base_get(b1) <
  cheri_base_get(b2) + cheri_length_get(b2)
);
\end{lstlisting}

\noindent
In practice, such a simple attack is unlikely to succeed on all but the
most basic allocators, as the most likely attack vector is when an allocator
fails to take into account bounds imprecision. The `full' \overlap
attack initially finds the first 512 lengths that are not precisely representable
as bounds and then randomly allocates multiple blocks to see if any of the
resulting capabilities overlap.

\subsection{Failed Attacks}

Two attacks in~\autoref{tab:atks} failed ($\oslash$) due to what we
believe are unintended bugs (as distinct from
\autoref{sec:escinauthentic}, where we could not distinguish unintended
bugs from careful, efficient, programming), and thus we cannot state whether the allocator is
vulnerable or invulnerable.

\escperms fails on \memalloc{bump-alloc-nocheri} because \fnc{realloc}
causes a \lstinline{SIGPROT} when trying to increase a block in size. By
design, \memalloc{bump-alloc-cheri} contains a version of \fnc{realloc} which
fixes this issue (see~\autoref{sec:adapting_to_cheri}).

\overlap fails on \memalloc{snmalloc-cheribuild} due to what we believe is
an internal snmalloc bug. Because of this, we included a newer version
of snmalloc as \memalloc{snmalloc-repo} which is invulnerable to \overlap.

\section{Performance Evaluation}
\label{sec:performance}

We wanted to understand both the relative performance of allocators under CHERI,
and also the performance impact of porting allocators to CHERI. The former
measures performance within hybrid and purecap and is
straightforward (\autoref{ssec:withinhybridpurecap}). The latter measures
performance across hybrid and purecap and shows far greater differences
than we expected (\autoref{ssec:withinhybridpurecap}). Because of that,
in \autoref{sec:dissection} we attempt to understand possible causes
for these differences.

\subsection{Methodology}

\begin{table}[t]
  \begin{center}
    \begin{tabular}{lll}
      \toprule
      Benchmark & Source & Characterisation \\
      \midrule
      barnes & mimalloc & Floating-point compute \\
      binary-tree & boehm & Alloc \& pointer indirection \\
      cfrac & mimalloc & Alloc \& int compute \\
      espresso & mimalloc & Alloc \& int compute \\
      glibc-simple & mimalloc & Alloc \\
      mstress & mimalloc & Alloc \\
      richards & richards & Pointer indirection \\
      \bottomrule
    \end{tabular}
  \end{center}
  \caption{Our benchmark suite. We list the benchmark name, the source
  of the benchmark, and a brief characterisation of it as a workload.
  \emph{Alloc} is short-hand for `allocator intensive' (i.e.~frequent
  allocation and deallocation).}
\label{tab:benchmarks}
\end{table}

\begin{figure*}[t]
  \begin{center}
    \includegraphics[width=\textwidth]{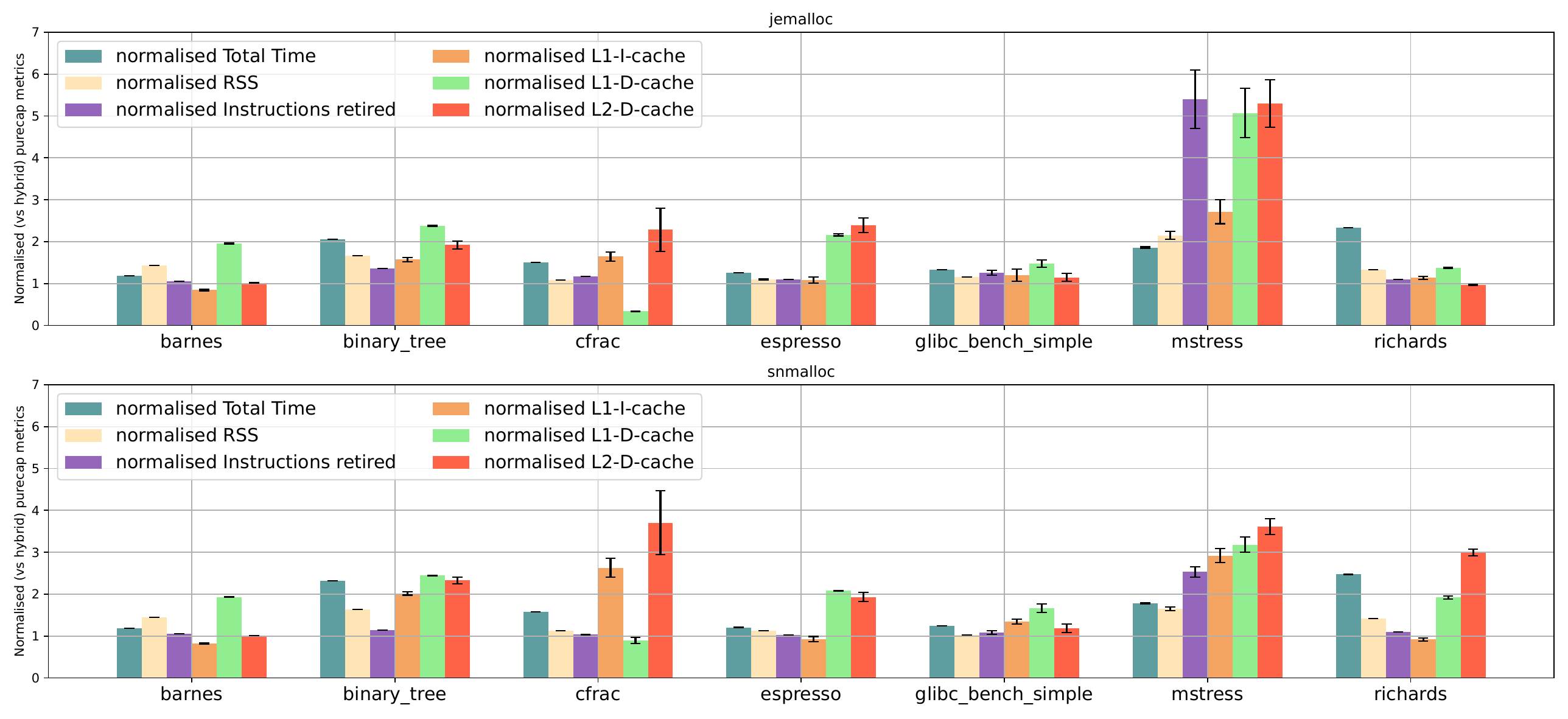}
    \vspace{-25pt}
  \end{center}
  \caption{\label{fig:eval jemalloc}
  \memalloc{jemalloc} (top) and \memalloc{snmalloc} (bottom) running on purecap, normalised
  to their respective hybrid allocators. For example, the wall-clock
  execution time \emph{total-time} of
  barnes with \memalloc{jemalloc} on purecap is 1.22x greater than
  on hybrid. To understand why purecap is slower than we expected, we recorded
  several performance metrics: \emph{rss-kb} is 
  maximum memory utilisation; \emph{INST\_RETIRED} the number of instructions
  retired while executing the benchmark; and \emph{\{L1-I, L1-D, L2-D\}\_CACHE}
  the L1 instruction, L1 data, and L2 data, cache misses respectively.
  None of these factors provides obvious clues as to why purecap is so
  much slower than hybrid.
  }
\end{figure*}

We conducted our experiments on Arm's prototype Morello hardware:
a quad-core Armv8-A 2.5GHz CPU, with 64KiB L1 data cache,
64KiB L1 instruction cache, 1MiB L2 unified cache per core, two 1MiB L3 unified
caches (each shared between a pair of cores), and 16GiB DDR4 RAM. We ran
CheriBSD 22.12 as the OS, with both purecap (\texttt{/usr/lib/}) and
hybrid (\texttt{/usr/lib64/}) userlands installed.

We wanted a benchmark suite that contains benchmarks written in C, that have
minimal library dependencies (so that we best understand what is being run),
and that execute fixed workloads (rather than those that execute for fixed
time). We selected 5 benchmarks from the \emph{mimalloc-bench}
suite~\cite{leijen19mimalloc} that meet this criteria, as well as the `classic'
binarytrees~\cite{boehm14artificial} and richards~\cite{richards99bench}
benchmarks. \autoref{tab:benchmarks} shows our complete benchmark suite.
Our benchmark suite deliberately contains a mix of allocation-heavy benchmarks
and non-allocation-heavy benchmarks, where the latter can serve as a
partial `control' to help us understand the effects of allocators on
performance against other factors.

We compile each benchmark with clang's \texttt{-O3} optimisation level. We
use \texttt{LD\_PRELOAD} at runtime to dynamically switch between allocators.
We measure wall-clock time on an otherwise unloaded Morello machine.
Since jemalloc and snmalloc were the highest performing allocators, we
concentrate on them for the rest of this section.

\subsection{Results Within Hybrid and Purecap}
\label{ssec:withinhybridpurecap}

With the geometric mean, snmalloc is faster than jemalloc by 1.25x
and 1.24x in hybrid and purecap respectively.
Overall, the wall-clock time roughly correlates with instruction counts
and L1I instruction cache misses. snmalloc retires a similar number of instructions
compared to jemalloc in hybrid and 1.19x fewer in purecap. snmalloc also
has 1.73x and 1.6x fewer L1-I cache misses than jemalloc in hybrid and purecap respectively.
It is difficult to know whether this difference is because snmalloc is
faster in general, or because it has been more carefully tuned for purecap than
has jemalloc --- indeed, a combination of both factors is plausible.
Because the overall performance differences are clear and because, as we will soon see, there are much
greater differences to consider elsewhere, we do not dwell further on these results.

\subsection{Results Across Hybrid and Purecap}

\autoref{fig:eval jemalloc} shows a comparison of \memalloc{jemalloc} and \memalloc{snmalloc}
across hybrid and purecap. We expected to see some performance difference between hybrid and purecap:
capabilities are likely to incur some costs due to their increased size
(e.g.~increased cache pressure); and purecap allocators make use of additional
security properties, such as bounds information, which require executing more
instructions. However, the differences are far greater than we expected:
using the geometric mean, jemalloc and snmalloc purecap
are 1.56x and 1.61x slower respectively than their hybrid counterparts.
This is a larger slowdown than we believe can be explained by the additional costs of using
capabilities in the allocator --- indeed, richards, a benchmark which performs
little allocation, also slows down by more than 2x.

\begin{figure}
  \begin{center}
    \includegraphics[width=1.0\columnwidth,trim={0.5cm 7cm 3.5cm 6.7cm},clip]{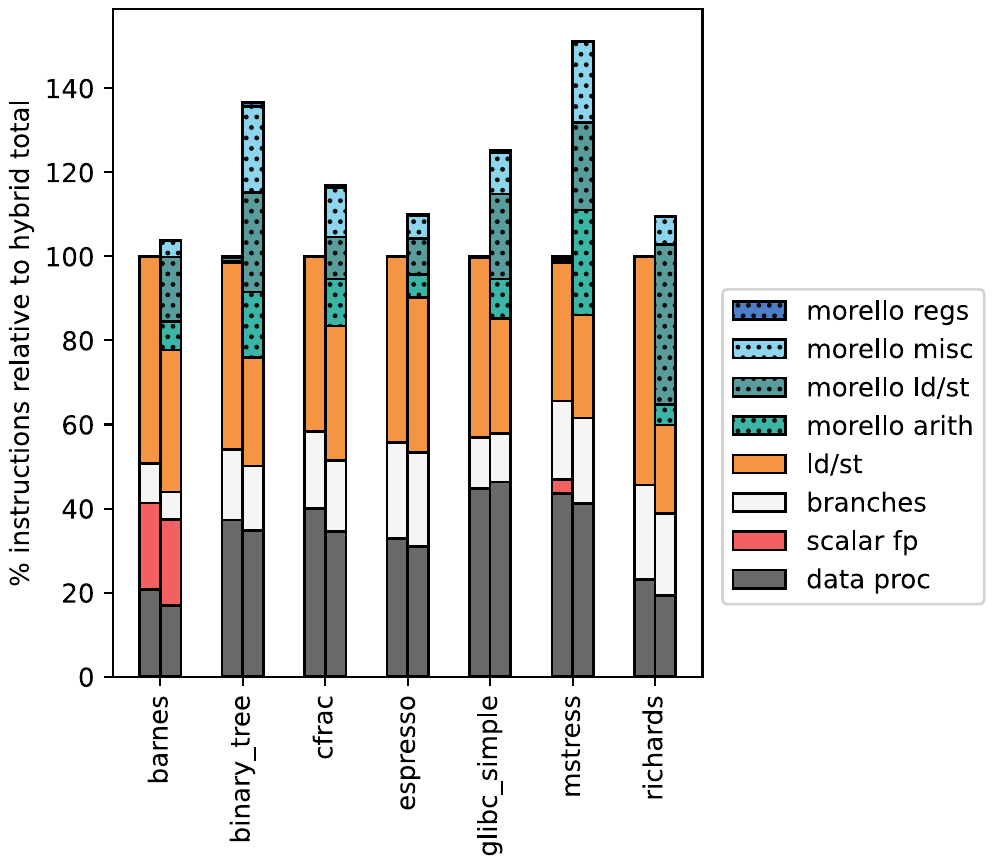}
    \vspace{-30pt}
  \end{center}
  \caption{\label{fig:instrmix}
  Dynamic instruction mix for hybrid (left-hand bar) and purecap (right-hand bar) benchmark runs.
  Most of these are as expected. For example, branches are the same in hybrid
  and purecap; some pointer arithmetic (a subset of `data proc' in hybrid)
  moves from AArch64 to Morello (`morello arith'). Similarly, some loads
  and stores move from AArch64 to Morello. There is a small but noticeable
  increase in the overall quantity of loads and stores (AArch64 and Morello
  combined). The `Morello misc' category captures instructions related to
  capability bounds, tag checks, and so on that we would only expect to see in
  any quantity in purecap.}
\end{figure}

\begin{figure}
  \begin{center}
    \includegraphics[width=1.0\columnwidth]{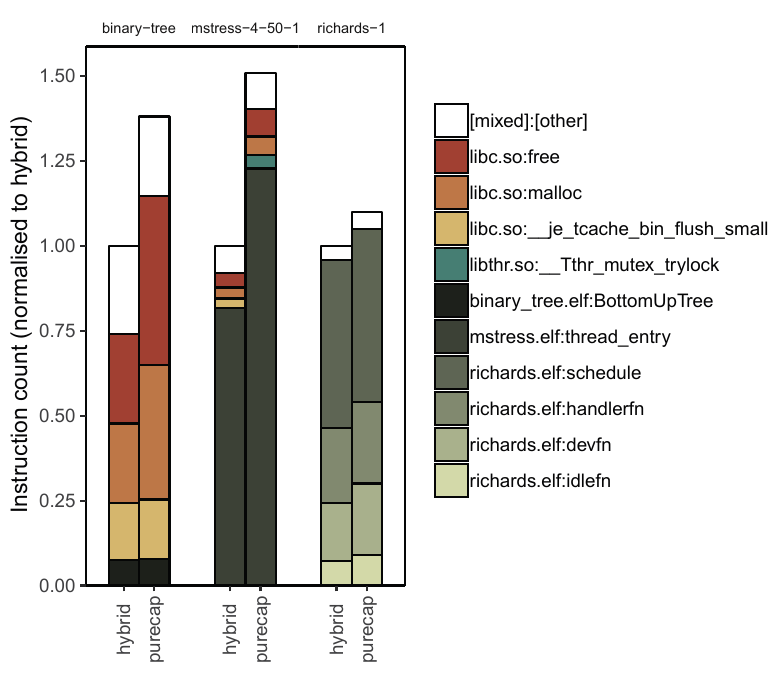}
    \vspace{-25pt}
  \end{center}
  \caption{\label{fig:fvpstats}
  Instruction counts per symbol, based on FVP traces for three benchmarks using
  the CheriBSD system allocator (jemalloc). mstress and richards were
  parameterised to perform fewer iterations than the default. Note that whilst
  binary\_trees is dominated by the effects of \fnc{malloc} and \fnc{free}, the others
  are dominated by their own code.
  }
\end{figure}

\section{Analysing The Disparity Between Hybrid and Purecap Performance}
\label{sec:dissection}

The disparity in performance between hybrid and purecap surprised us,
and we thus explored three different avenues in an attempt to understand
the causes. In this section, we explain these avenues in ascending order of
difficulty.

Our results suggest that it is likely that much of the difference we see is not inherent to the use
of capabilities in a CPU, but is the result of immature tool-chains and
micro-architectural anomalies common in a first prototype implementation of an
ambitious platform such as Morello. Our results should be interpreted in
that context: it is likely that further research, and future evolutions of
CHERI hardware, will change our view of capabilities' effect on performance.
Indeed, since we submitted this paper we have become aware of other researchers who are also
investigating similar issues\footnote{Although not available at the time of writing, we have been
informed that a detailed performance white paper will soon be available at
\url{https://www.arm.com/-/media/Files/pdf/white-paper/prototype-morello-performance-results}}.

\subsection{Simple Metrics}

We started by measuring several simple metrics, including the
resident set size (RSS), and several CPU performance counters: the results are
shown in ~\autoref{fig:eval jemalloc}.
We hoped that this might highlight possible causes such as increased cache pressure
in purecap. Unsurprisingly, overall purecap has higher costs across all
metrics. L2 data cache accesses increase in almost all cases in purecap,
suggesting increased pressure on the L1 data cache with capabilities. In most
cases memory usage is only slightly worse in purecap, though in the
allocator-intensive mstress RSS doubles.
However, overall, these factors do not suggest an obvious cause for the slowdown
we see.

\subsection{Class, and Quantity of, Instructions}

We wondered whether the higher instruction counts on purecap benchmarks
could be explained by the extra CHERI instructions that
an allocator needs (compare Listings~\ref{lst:bump_alloc1} and~\ref{lst:bump_alloc2}).
To understand this, we wanted
to count how often different `classes' of instructions were executed.
Unfortunately, there is currently no direct way on either CheriBSD or Morello
to obtain such a count, so we had to cobble together two approaches: a fast,
somewhat imprecise, approach based on QEMU; and a slow, but more
precise, approach based on Arm's Morello Platform FVP~\cite[p.~23]{morellofvp}.
In essence, the latter serves as a sanity check for the former.

We wrote a custom QEMU plugin that maintains counts for each class of
instructions as a Morello instance is running. This includes the kernel booting
and shutting down as well as the benchmark we are interested in. We repeatedly
ran CheriBSD without any meaningful workload, so that we could find the
instruction counts that together constitute the `head' and `tail' of execution.
When we ran a benchmark, we then subtracted the head/tail instruction counts to
obtain the `benchmark only' instruction counts.

To validate these results, we used the Morello Platform FVP, which can emit \emph{tarmac}
traces~\cite[p.~5452]{fvpguide} representing complete records of execution. We
altered the FVP so that when we executed an otherwise unused instruction, it
toggled tracing on and off. We executed a complete run of the binary\_tree
benchmark, examined the traces, and counted the instructions contained therein,
which were a close match to our QEMU instruction counts. This gives us confidence
that our QEMU figures are representative.
Since our QEMU approach only has to track a few integers, whereas FVP produces
tarmac traces that are often a TiB long for our workloads, our two approaches
differ in performance by about 3 orders of magnitude. Running our full
benchmark suite for its full duration would be infeasible in FVP, so the
results are from our QEMU approach.

The instruction mixes in \autoref{fig:instrmix} look largely sensible: some
aspects (e.g.~branches) are identical in hybrid and purecap; some
vary where capabilities are sometimes used (e.g.~loads and stores);
and purecap uses Morello instructions. Overall, while there are some
minor oddities (e.g.~mstress uses Neon vector instructions -- classified as
floating point -- to a much greater degree on hybrid compared to purecap), there are no obvious smoking
guns.

We then wrote an analysis tool for our FVP traces (which contain virtual
addresses) to provide per-function profiling, and ran this for one full
benchmark, and shortened versions of two others.
\autoref{fig:fvpstats} shows that some
functions execute many more instructions in purecap (vs.~hybrid) than
others. \fnc{malloc} and \fnc{free} are particularly strongly affected, but the
effect on other functions is fairly uniform.

\subsection{Low-level Differences}

Hybrid and purecap CHERI imply certain differences which we would expect to
account for some of the performance changes we see. In this
section we analyse the following factors in detail:

\newcommand\fhardware{F$_{\textrm{Hardware}}$\xspace}
\newcommand\fabi{F$_{\textrm{ABI}}$\xspace}
\newcommand\fcompiler{F$_{\textrm{Toolchain}}$\xspace}
\newcommand\fuser{F$_{\textrm{User}}$\xspace}

\setlength{\leftskip}{6pt}

\vspace{6pt}\noindent\textbf{\fhardware}
At the hardware level, pointer operations (including arithmetic, loads, and
stores) have different semantics for capabilities, which are likely to have
different performance characteristics relative to operations on normal
pointers. Similarly, since capabilities are double word width, we would expect
them to put greater pressure on caches and other system resources.

\vspace{6pt}\noindent\textbf{\fabi}
At the ABI level, the purecap CheriBSD ABI is different than the hybrid ABI
(where the latter is largely the same as a non-CHERI ABI), sometimes
requiring different code generation.

\vspace{6pt}\noindent\textbf{\fcompiler}
At the compiler level, both hybrid and purecap modes require altered versions
of LLVM. Neither is as mature as ``mainstream'' LLVM, and thus are unlikely to
optimise code as fully as expected. Although
the purecap LLVM has received more attention than the hybrid LLVM, it is also
more different from ``mainstream'' LLVM, so it is possible that the purecap
LLVM will produce less optimal code than the hybrid LLVM.

\vspace{6pt}\noindent\textbf{\fuser}
At the user level, code that wants to take advantage of CHERI will tend to use
different execution paths when compiled for purecap (e.g.~the bump
allocator of~\autoref{lst:bump_alloc1}).

\setlength{\leftskip}{0pt}

\vspace{6pt}\noindent
Our expectation is not that we can identify causal relationships
for performance oddities, but that we can at least understand some of
the `beneath the surface' factors that might explain part of the performance
story.

\subsubsection{\fhardware: Hardware pointer operation (micro\-benchmarks)}

\begin{figure}
  \begin{center}
    \includegraphics[width=\columnwidth]{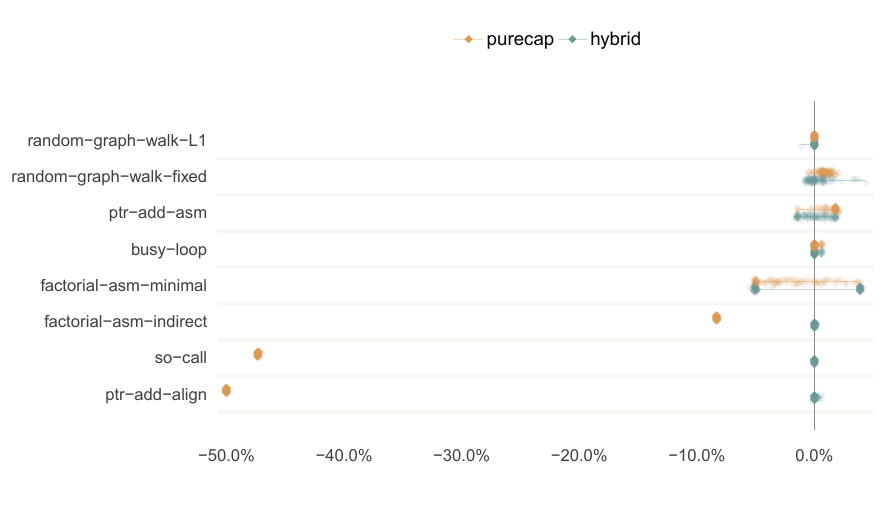}
    \vspace{-30pt}
  \end{center}
  \caption{\label{fig:mbench}
  Microbenchmark performance results, with purecap normalised to median hybrid
  performance (slower to the left, faster to the right). This shows that many
  operations have similar performance on hybrid and purecap but some, such as
  calling a function in a shared object, are significantly slower on purecap. }
\end{figure}

To attempt to understand some of the performance characteristics of Morello, we wrote a
series of simple microbenchmarks, in a variety of C and assembly, that run
under both hybrid and purecap CheriBSD. The results are shown in \autoref{fig:mbench}.

We split our microbenchmarks into two. First, those microbenchmarks which
show little or no difference between hybrid and purecap:

\setlength{\leftskip}{6pt}

\vspace{6pt}\noindent
\emph{random-graph-walk-L1} performs a random walk through a
    graph that fits in Morello's 64KiB L1 data cache. This
    suggests that there is no overhead in reading from a capability.

\vspace{6pt}\noindent
\emph{random-graph-walk-fixed} is similar, but uses a larger set
    that consumes 1MiB in hybrid and (due to capabilities
    being double word width) 2MiB in purecap. Despite this, there is little
    performance difference between hybrid and purecap.

\vspace{6pt}\noindent
\emph{ptr-add-asm} measures pointer addition, a common operation.
    Although capability addition is more complex (due to bounds checking),
    it has a dedicated instruction and hybrid and purecap perform similarly.

\vspace{6pt}\noindent
\emph{busy-loop} is an empty C loop, to act as a control. The C
    loop uses no capabilities, and compiles to identical code for hybrid and
    purecap targets. Hybrid and purecap have the same performance.

\setlength{\leftskip}{0pt}

\vspace{6pt}\noindent
Second, those benchmarks which do show noticeable differences:

\setlength{\leftskip}{6pt}

\vspace{6pt}\noindent
\emph{factorial-asm-minimal} is a tail-recursive factorial
    implementation that, despite not using capabilities in the loop, shows
    bimodality in hybrid. We assume this is a microarchitectural artefact.

\vspace{6pt}\noindent
\emph{factorial-asm-indirect} is similar, but uses an indirect
    tail call of a normal pointer in hybrid and a capability in purecap.
    This benchmark shows a consistent overhead of just under
    10\% in calling via a capability.

\vspace{6pt}\noindent
\emph{so-call} measures the PLT overhead by having a simple loop
    call an empty function in another shared object. This operation
    is significantly slower in purecap. We assume this is a
    microarchitectural artifact.

\vspace{6pt}\noindent
\emph{ptr-add-align} performs pointer addition followed by an
    align-down operation, a common operation in allocators. The capability
    variant has to check bounds encodability and so uses a dedicated
    instruction, while the hybrid version uses a simple bitwise operation. As a
    result, the purecap version is much slower than the hybrid version.

\setlength{\leftskip}{0pt}

\vspace{6pt}\noindent
Since the \emph{so-call} and \emph{ptr-add-align} microbenchmarks show
significant differences between purecap and hybrid, we analysed how often the
same idioms occurred in our larger benchmarks. In the three benchmarks shown in
\autoref{fig:fvpstats}, these two factors account for 0.00\%-1.25\% of executed
instructions (\autoref{tab:fvp_insn_report}) --- they are thus likely to play a
correspondingly small part in
the performance difference seen in \autoref{fig:eval jemalloc}.

\subsubsection{\fabi: ABI Implications}
\label{ssec:purecap ABI difficulties}

We observed some ABI-related differences between hybrid and purecap code.
For example, storing zero to a global variable in hybrid compiles
to the following:

\begin{lstlisting}
    adrp x1, #+0x20000
    str xzr, [x1, #2472]
\end{lstlisting}

\noindent However, the purecap ABI
requires another level of indirection in order to obtain a capability with
tight bounds:

\begin{lstlisting}
    adrp c1, #+0x10000
    ldr c1, [c1, #3776]
    str xzr, [c1]
\end{lstlisting}

\noindent
It is difficult for us to tell if this behaviour is important for
security in all, or merely some, situations. However, analysing
the traces behind~\autoref{fig:fvpstats} showed that
in the richards benchmark,
for example, we were able to observe that such code was executed
frequently in some functions, though we cannot precisely quantify
the performance impact.

\subsubsection{\fcompiler: Toolchain maturity}

We noticed that the same C code compiled for hybrid and purecap by LLVM could
sometimes result in very different machine code. The differences are
far too extensive to admit a simple analysis, but we hypothesise that
they might be due to capability code: either preventing LLVM from performing
some of its normal optimisations when capability code is present; or LLVM
simply not having been taught how to optimise capability code. Two small examples
demonstrate the overall point.

In hybrid code, zeroing memory is compiled to a single instruction:

\begin{lstlisting}
stp     xzr, xzr, [x0]
\end{lstlisting}

\noindent whereas in purecap it is compiled to two instructions:

\begin{lstlisting}
movi    v0.2d, #0000000000000000
stp     q0, q0, [c0]
\end{lstlisting}

\noindent The purecap version could have used a single instruction
with the \texttt{czr} register. This may or may not improve performance,
but serves as an example of how quickly minor code generation differences
can make it difficult for humans to comprehend differences between
hybrid and purecap code generation.

Another example relevant to allocators is in a hot path
within jemalloc's \fnc{malloc} function, where a byte-sized th\-read-local
is loaded from memory. In hybrid this is compiled to:

\begin{lstlisting}
ldr     x2, [...]
mrs     x1, TPIDR_EL0
ldrb    w0, [x2, x1]
\end{lstlisting}

\noindent
In purecap this is compiled to a load and a bounds restriction:

\begin{lstlisting}
ldp     x2, x3, [...]
mrs     c1, CTPIDR_EL0
add     c2, c1, x2
scbnds  c2, c2, x3
ldrb    w0, [c2]
\end{lstlisting}

\noindent
These additional instructions will almost certainly have a measurable impact
on performance, but do not appear to have any security benefit:
the \texttt{c2} register is not indexed by a variable, and a greater capability is
already available in \texttt{c1}. It seems likely that this is a missed
optimisation opportunity by the compiler rather than a deliberate security restriction.

\subsubsection{\fuser: The Impact of CHERI Security on User Code}

CHERI aware allocators often improve security by restricting capability bounds
and other permissions. Doing so requires generating and running more code.
For example, after allocating space internally, a purecap \fnc{malloc}
may need to derive a capability from a `super' capability:

\begin{lstlisting}
...     c0, ...  # Calculate address
...     x1, ...  # Calculate length
scbndse c0, c0, x1
mov     x2, #0xffffffffffff....
movk    x2, #0x...., lsl #16
clrperm c0, c0, x2
\end{lstlisting}

\noindent
In this fragment, \texttt{scbndse} sets the bounds, and \texttt{clrperm} removes
permissions that \fnc{malloc} results should not have. Although it is
difficult for us to say with certainty, it appears that such instructions
are not as well optimised as they could be: we observed situations where
it would seem more efficient to reorder some of these instructions,
exposing further opportunities for optimisation. Either way, it would
be interesting to understand the performance impact of these CHERI-aware
aspects in isolation from other aspects of code generation, but this
would require a very challenging analysis that is beyond the scope of our work.

\section{Conclusions}

CHERI holds great promise for securing software in general: allocators are a
key part of that story. In this paper we have shown that many CHERI allocators,
including the current CheriBSD default allocator, suffer from simple security
vulnerabilities. We have also shown that measuring the effect of
capabilities on performance is challenging, as it is difficult to understand, let
alone factor out, the impact of immature compiler toolchains and prototype
hardware. We expect ongoing and future research to tease apart these factors.

Despite all of this, one allocator has shone throughout: snmalloc is
not susceptible to any of our attacks, and is faster than the default CheriBSD
allocator in our benchmarks. We suggest that snmalloc be considered to be the
default CheriBSD allocator going forward.

\textbf{Acknowledgements:} We thank Ruben Ayrapetyan, David Chisnall,
Jessica Clarke, and Richard Grisenthwaite for comments. This work was funded by the
Digital Security by Design (DSbD) Programme delivered by UKRI
(including grants EP/V000349/1 and EP/V000373/1).

\begin{table*}[htp]
  \begin{center}
    \begin{tabular}{lccccc}
      \toprule
      Name & \multicolumn{5}{c}{\% of instructions}\\
      \cmidrule(lr){2-6}
        & \texttt{.plt} entries & \texttt{ALIGND} & \texttt{BLR <cap>} & \texttt{BR <cap>} & \texttt{RET <cap>}\\
      \midrule
      hybrid-binary-tree & 0.92 & 0.00 & 0.00 & 0.00 & 0.00\\
      purecap-binary-tree & 0.68 & 0.57 & 0.00 & 0.68 & 1.48\\[6pt]
      hybrid-mstress & 0.24 & 0.00 & 0.00 & 0.00 & 0.00\\
      purecap-mstress & 0.18 & 0.13 & 0.00 & 0.18 & 0.52\\[6pt]
      hybrid-richards & 0.00 & 0.00 & 0.00 & 0.00 & 0.00\\
      purecap-richards & 0.00 & 0.00 & 1.36 & 0.00 & 1.36\\
      \bottomrule
    \end{tabular}
  \end{center}
  \caption{Detailed instruction statistics (as a percentage of all instructions executed) for the benchmarks presented in
  \autoref{fig:fvpstats}. As expected, none of the benchmarks used branch-to-capability
  in hybrid. In purecap, `\lstinline{.plt} entries', which counts the number of times a \lstinline{.plt}
  section was entered (by any means), is virtually the same as `\lstinline{BR <cap>}'
  because each purecap PLT entry ends with \lstinline{BR c17}.
  `\lstinline{.plt} entries' and `\lstinline{ALIGND}' correspond to
  the \emph{so-call} and \emph{ptr-add-align} micro\-benchmarks, respectively,
  showing that any performance degradation from these micro-benchmarks can
  explain only a small part of the overall slowdowns we see.
  We believe that the `\lstinline{BLR <cap>}' results for purecap-richards are calls
  via function pointers, and that these have performance characteristics similar
  to \lstinline{BR <cap>} and thus to the \emph{so-call} micro-benchmark. In other words,
  again, these are likely to explain only a small part of the overall slowdowns we see.}
\label{tab:fvp_insn_report}
\end{table*}

\bibliographystyle{ACM-Reference-Format}
\bibliography{bib}

\end{document}